\documentclass{ws-procs10x7}
\usepackage{balance}
\usepackage{epsfig}

\newcommand{\qq}{\mbox{$\mathrm{q} \bar{\mathrm{q}}$}}
\newcommand{\ff}{\mbox{$\mathrm{f} \bar{\mathrm{f}}$}}

\newcommand{\bb}{\mbox{$\mathrm{b} \bar{\mathrm{b}}$}}

\newcommand{\mZ}{\mbox{$M_{\mathrm{Z}}$}}

\newcommand{\mh}{\mbox{$m_{\mathrm{h}}$}}
\newcommand{\mA}{\mbox{$m_{\mathrm{A}}$}}

\newcommand{\mW}{\mbox{$m_{\mathrm{W}^{\pm}}$}}

\newcommand{\tanb}{\mbox{$\tan \beta$}}

\newcommand{\msusy}{\mbox{$M_{\rm SUSY}$}}

\newcommand{\Mx} {\mbox{$ m_{\mathrm H_1} \, $}}

\newcommand{\Hx} {\mbox{$ {\mathrm H_1} \, $}}
\newcommand{\Hy} {\mbox{$ {\mathrm H_2} \, $}}
\newcommand{\Hz} {\mbox{$ {\mathrm H_3} \, $}}

\newcolumntype{d}[1]{D{.}{.}{#1}}

\makeindex
\begin{document}

\title{RESULTS OF BEYOND THE STANDARD MODEL HIGGS SEARCHES FROM THE LEP EXPERIMENTS}

\author{A. LUDWIG$^*$ (ON BEHALF OF THE FOUR LEP COLLABORATIONS)}

\address{Physikalisches Institut, Universit\"at Bonn, Nussallee 12, 53115 Bonn, Germany\\$^*$E-mail: aludwig@physik.uni-bonn.de }


\twocolumn[\maketitle\abstract{Since no Standard Model Higgs was discovered at LEP the searches were extended to beyond the Standard Model Higgs scenarios. 
A selection of final results from searches carried out by the four LEP experiments ALEPH, DELPHI, L3 and OPAL are presented that
 include data taken at the highest centre-of-mass energies.}
]
\section{Introduction}

Searches exploring the Higgs sector of beyond the Standard Model (bSM) scenarios are reported in the following submitted papers.
The LEP Collaborations have combined their final results of searches within the constrained minimal supersymmetric Standard Model (cMSSM) including CP conserving and CP violating scenarios\cite{unknown:2006cr}. L3 finalised their search for a narrow invisibly decaying Higgs\cite{Achard:2004cf}. For the first time OPAL performed a search for a broad invisibly decaying Higgs\cite{mydraft}. Finally L3 updated a search for anomalous couplings of the Higgs\cite{Achard:2004kn}.

\section{Searches in the cMSSM}
In the cMSSM, at tree level, two parameters are sufficient to fully describe the Higgs sector. A convenient choice is one Higgs mass
 and the ratio $\tanb=v_2/v_1$ of the vacuum expectation values of the two Higgs fields.
 The values of five SUSY breaking parameters and two phases that occur in radiative corrections are fixed at the electroweak scale in so-called benchmark scenarios\cite{unknown:2006cr}. The predictions of each benchmark model depend strongly on the measured top quark mass $m_{\rm top}$.
At LEP the cMSSM Higgs can be produced in Higgsstrahlung and in pair production. For a very light Higgs Yukawa production is also relevant. The complementarity of these main production processes, (see Table\,\ref{tab1}), ensures high sensitivity over the full accessible cMSSM parameter space. 
The channels investigated in these searches are tabulated in Table\,\ref{tab0}. In the following two examples of the set of investigated scenarios will be highlighted.
\begin{table}
\vspace{-4.0mm}
\tbl{Cross-sections in the cMSSM expressed in terms of the SM Higgs production cross-section $\sigma_{\rm{HZ}}^{\rm SM}$. Here $\alpha$  is the mixing angle which diagonalises the mass matrix of the CP-even Higgs and $\bar\lambda$ is a kinematic factor (see Ref. 1).\label{tab1}}
{\begin{tabular}{@{}c c@{}}
\toprule
Higgsstrahlung & Pair Production\\ \colrule
$\sigma_{{\rm h}{\rm Z}} = \sin^2(\beta - \alpha)~\sigma_{\rm{HZ}}^{\rm SM}$
 &
$\sigma_{{\rm h}{\rm A}} = \cos^2(\beta - \alpha){\bar \lambda}~\sigma_{\rm{HZ}}^{\rm SM}$
\vspace{0.3cm}  \\ 
$\sigma_{{\rm H}{\rm Z}} = \cos^2(\beta - \alpha)~\sigma_{\rm{HZ}}^{\rm SM}$
&
$\sigma_{{\rm H}{\rm A}} = \sin^2(\beta - \alpha){\bar \lambda}~\sigma_{\rm{HZ}}^{\rm SM}$
  \\
\botrule
\vspace{-8.0mm}
\end{tabular}}
\end{table}

\vspace{-4.0mm}
\begin{table}
\tbl{ Searches for \Hx (\Hy), the (second-) lightest neutral cMSSM Higgs.\label{tab0}}
{\begin{tabular}{@{}l c l@{}}
\toprule
Production & $\otimes$ & Decays                                           \\ \colrule

$e^+e^- \to \mathrm{ZH_1~or~ZH_2}$ & $$ & $\mathrm{ H_1}  \to \mathrm{b} \bar{\mathrm{b}} ,\mathrm{q} \bar{\mathrm{q}}, \tau^+\tau^- $  \\ 
$e^+e^- \to \mathrm{ H_1H_2}    $ & $$ & $\mathrm {H_2}\to \mathrm {H_1H_1},\mathrm{b} \bar{\mathrm{b}}, \tau^+\tau^- $                  \\
$e^+e^- \to \mathrm{b} \bar{\mathrm{b}}\mathrm{H_1}$ &$$ &                                           \\ \botrule
\end{tabular}}
\vspace{-7.0mm}
\end{table}

\subsection{\it CP conserving scenarios}
The {\it \mh-max} scenario\cite{unknown:2006cr} is designed to maximise the theoretical upper bound on the lightest Higgs mass \mh\ for a given \tanb , fixing $m_{\rm top}$
and the soft SUSY-breaking mass parameter \msusy\ , by setting the stop mixing parameter $ X_{\rm t}$=2 \msusy\ to 2\,TeV. Thus this model provides the most conservative exclusion limits for \tanb.  
\begin{figure}[htbp]
  \vspace{-9.0mm}
     \centerline{\epsfig{figure=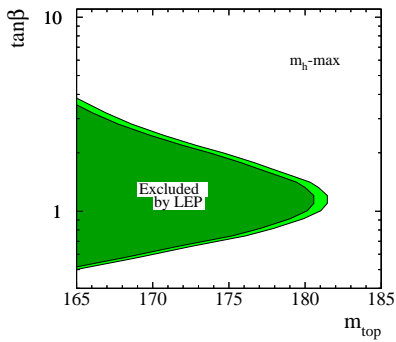,width=2.2in}}
  \vspace{-2.0mm}
  \caption{Domains of \tanb\ which are excluded at the 95\% CL (light shaded) and the 99.7\% CL (dark shaded), 
    for the CP-conserving {\it \mh-max} benchmark scenario, as a function of the assumed top quark mass.}
  \label{fig2}
  \vspace{-4.0mm}
\end{figure}
The observed exclusion at 95\% CL of \mh\ up to 92.8\,GeV (expected 94.9\,GeV) and \mA\ up to 93.4\,GeV (expected 95.2\,GeV) is more or less insensitive to $m_{\rm top}$. However the exclusion of \tanb\ is sensitive to $m_{\rm top}$ and may vanish completely for $m_{\rm top}> 181$\,GeV (Fig.\,\ref{fig2}).
\subsection{\it CP violating scenarios}
\begin{figure}[htbp]
  \begin{center}
\vspace{-9.0mm}
   \includegraphics[width=2.2in]{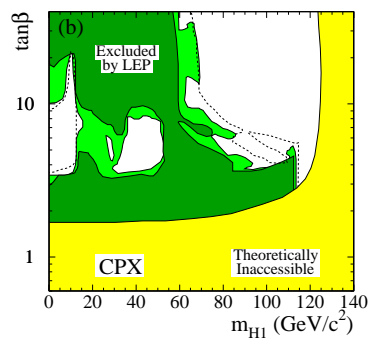} 
  \end{center}
\vspace{-5.0mm}
  \caption{Regions which are excluded at the 95\% CL (light shaded) and the 99.7\% CL (dark shaded) in the \tanb\ versus \Mx plane. The expected exclusion is delimited by the dashed contours.}
  \label{fig3}
\vspace{-4.0mm}
 \end{figure}

The interpretation in CP violating scenarios is currently based only on data taken by the OPAL detector\cite{Ahmet:1990eg}. In such scenarios the three neutral Higgs mass eigenstates are mixtures of CP-even and CP-odd fields. While cascaded decays may be enhanced in some model points, leading to a good discovery potential, the absence of the \Hx ZZ coupling and a kinematical suppression of the \Hy\ and \Hz\ production in other model points makes the experimental search more challenging.
In the so-called CPX scenario\cite{unknown:2006cr} a substantial CP violation that may account for the observed cosmic matter-antimatter asymmetry can be induced by complex  phases in the soft SUSY-breaking sector. This gives rise to CP-even/odd mixing, proportional to $\frac{m^4_{\rm top}~\mathrm{Im}(\mu A)}{(v_1^2 + v_2^2)~M^2_{\mathrm{SUSY}}}.$
~Therefore, \msusy\ , the common trilinear Higgs-squark coupling A and the ``Higgs mass mixing parameter" $\mu$ are set to 500\,GeV, 1\,TeV and 2\,TeV, respectively.
Large Im($\mu A$), are obtained if the CP-violating phase $\arg (A)$ takes values close to $90^\circ$. 
Again, the effects from CP violation strongly depend on the precise value of $m_{\rm top}$.
Fig.\,\ref{fig3} shows excluded regions in the \tanb\ versus \Mx plane for $m_{\rm top}$=174.3\,GeV. 

A hole growing with the value of $m_{\rm top}$ occurs in the intermediate \tanb\ region, preventing an absolute mass limit on the lightest neutral scalar to be set. Very small values of \tanb\ are not preferred in the CPX scenario and can be excluded up to 2.9 at 95\% CL.  
\section{Searches for an invisible Higgs}
Invisible Higgs decays, predicted by some bSM scenarios, would lead to a characteristic signature in the LEP detectors. Two acoplanar jets or leptons with an invariant mass close to \mZ,~balancing a large transverse missing momentum and energy $E_{\rm miss}$ from the Higgs. In case of a narrow Higgs the recoil mass spectrum will peak at \mh. 
\subsection{\it Higgs with narrow width}
\begin{figure}[htbp]
\vspace{-6.0mm}
   \centerline{\epsfig{figure=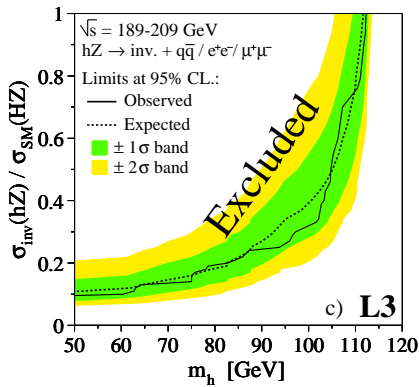,width=2.2in}}
\vspace{-2.0mm}
   \caption{Upper limits on the cross-section for the production of an invisibly decaying Higgs.}
   \label{fig4}
\vspace{-3.0mm}
 \end{figure}

 Higgs decays into graviscalars, Majorons or the lightest supersymmetric particle would be invisible but the decay-width would not be experimentally resolvable. 
L3 analysed 630 pb$^{-1}$ of data taken at centre-of-mass energies $\sqrt{s}>$189\,GeV in the channels  $e^+e^- \to {\rm ZH}\to \ff + E_{\rm miss},~({\rm f}\in\{\mu^-,e^-,{\rm q}\})$. Fig.\,\ref{fig4} shows the excluded branching ratio times production cross-section normalised to the SM cross-section.
For an assumed SM production rate one reads off immediately the excluded branching ratio of Higgs into invisible. E.g. a branching $>50\%$~is excluded up to $\mh\approx $ 105\,GeV.
If the invisible branching ratio of the Higgs would be 100\%, a lower mass limit \mh~of 112.3\,GeV (expected 111.6\,GeV) is set. This limit can be compared with previously published results\cite{Heister:2001kr,Abdallah:2003ry} reporting a lower limit on \mh\ of 114.1\,GeV (expected 112.6\,GeV) and \mh=112.1\,GeV (expected 110.5\,GeV), respectively.

\subsection{\it Higgs with large width}
\begin{figure}[htbp]
\vspace{-4.0mm}
  \centerline{\epsfig{figure=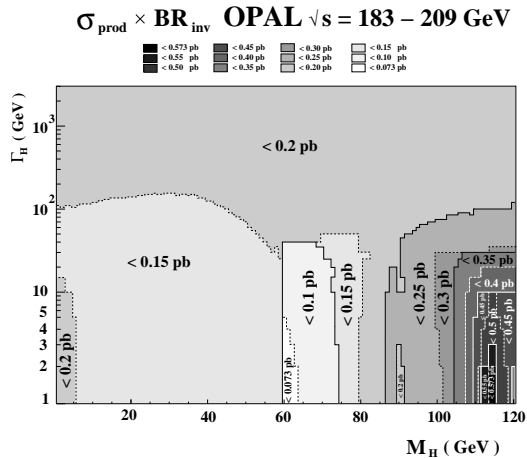,width=2.7in}}
\vspace{-2.0mm}
  \caption{Upper limits on the production cross-section times branching ratio of an invisibly decaying Higgs with decay-width $\Gamma_{\mathrm{H}}$.}
  \label{fig5}
\vspace{-6.0mm}
 \end{figure}

\begin{figure}[htbp]
\vspace{-4.0mm}
 \centerline{\epsfig{figure=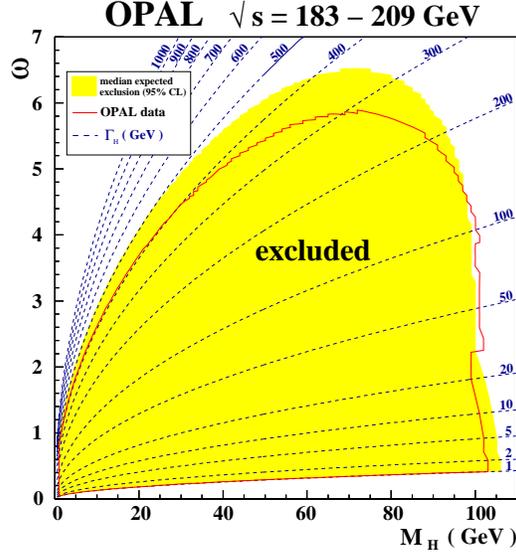,width=2.7in}}
\vspace{-3.0mm}
 \caption{Regions excluded at 95\% CL for a stealthy Higgs with mass $\mathrm{M_H}$ and decay-width $\Gamma_{\mathrm{H}}$, interpreted in terms of the coupling $\omega$.}
 \label{fig6}
\vspace{-4.0mm}
 \end{figure}

 Models with large extra dimensions\cite{mydraft} or extra singlet models\cite{mydraft} allow invisible decays of the Higgs with a decay-width $\Gamma_{\mathrm{H}}$ much larger than the experimental resolution even for a light Higgs. OPAL performed a model independent search in the parameter range 1\,GeV$<\mh<$120\,GeV and 1\,GeV$<\Gamma_{\mathrm{H}}<$3\,TeV in the channel $e^+e^- \to {\rm ZH(\mh,\Gamma_{\rm H})}\to \qq + E_{\rm miss} $ using data corresponding to an integrated luminosity of 630\,pb$^{-1}$ taken at $\sqrt{s} >$183\,GeV. 
The upper limits set on production cross-section times branching ratio are displayed in Fig.\,\ref{fig5}. In most of the search plane upper limits of the order of 0.1 to 0.2\,pb could be set. For extremely large Higgs widths of several hundred GeV, these limits become almost constant.  
 An interpretation of the search for an invisibly decaying Higgs with large $\Gamma_{\mathrm{H}}$ in terms of the coupling $\omega$ of the stealthy Higgs scenario is shown in Fig.\,\ref{fig6}. 
In the stealthy Higgs scenario\cite {mydraft} the Higgs couples to a hidden scalar sector via a non perturbative coupling $\omega$.
Since these scalars occur only in two-loop corrections, they are not excluded by electroweak precision measurements. If the coupling or the number of additional scalars is large they provide invisible decay channels for the Higgs.
Couplings $\omega$ up to 5.9 (\mh=73\,GeV) and corresponding widths from about $\Gamma_{\rm H} \approx$115\,GeV (at \mh=100\,GeV) up to $\Gamma_{\rm H} \approx $400\,GeV (for \mh $\lesssim$40 \,GeV) could be excluded. 
Very small couplings corresponding to $\Gamma_{\rm H}< 1$\,GeV can be excluded\cite{Abbiendi:2002qp} up to \mh=81\,GeV.    

 \section{Anomalous Higgs couplings}
 General effective theories may contain anomalous couplings (AC) for vertices of the type ${\rm HZ}\gamma$ and ${\rm H}\gamma\gamma$ at Born level. 
 Using the channels listed in Table\,\ref{tab2}, L3 probed \mh~between 70\,GeV up to 190\,GeV in a data set corresponding to an integrated luminosity of 602\,pb$^{-1}$ taken at $\sqrt{s}>$189\,GeV. The couplings are expressed in dimensionless parameters that vanish in the SM. Results for these parameters are complementary to that from triple gauge coupling analyses\cite{Achard:2004kn}. Only the measurements of d (see Fig.\,\ref{fig7}),the AC of the Higgs to the $\rm W^{1,2,3}$ gauge bosons, and of ${\rm d_b}$, the AC of the Higgs to the B boson, are competitive. Note that in the neighbourhood of d=0 the SM Higgs searches contribute to the limit and for $\mh>2\mW$~the sensitivity drops rapidly.
Taking the limits on d and ${\rm d_b}$ as inputs, two-dimensional limits on BR(${\rm H \to Z \gamma}$) versus BR(${\rm H \to \gamma \gamma}$) can be set, which exclude branching ratios of larger than 50\% to 60\% for \mh$>$120\,GeV. The search is not sensitive to the SM prediction of BR({$\rm H \to \gamma\gamma) \approx 2\times 10 ^{-3} $ at \mh=120\,GeV and BR({$\rm H \to Z \gamma) \approx 3\times 10 ^{-3} $ at \mh= 140\,GeV.
     
     \begin{table}
       \vspace{-5.0mm}
       \tbl{Channels sensitive to AC\label{tab2}.}
	   {\begin{tabular}{@{}l l@{}}
	       \toprule
	       Production & Decays\\ \colrule
	       $e^+e^- \to {\rm ZH}$ & ${\rm H}\to \bb ,\qq ,\gamma\gamma $ \\ 
	       $e^+e^- \to {\rm H}\gamma$ & ${\rm H}\to {\rm Z}\gamma, {\rm WW^*}, \gamma\gamma $\\
	       $e^+e^- \to {\rm He^+e^-}$ & ${\rm H}\to  \gamma\gamma $ \\
	       \botrule
	   \end{tabular}}
     \end{table}

     \begin{figure}[htbp]
       \vspace{-8.0mm}
       \begin{center}
	 \includegraphics[width=2.2in]{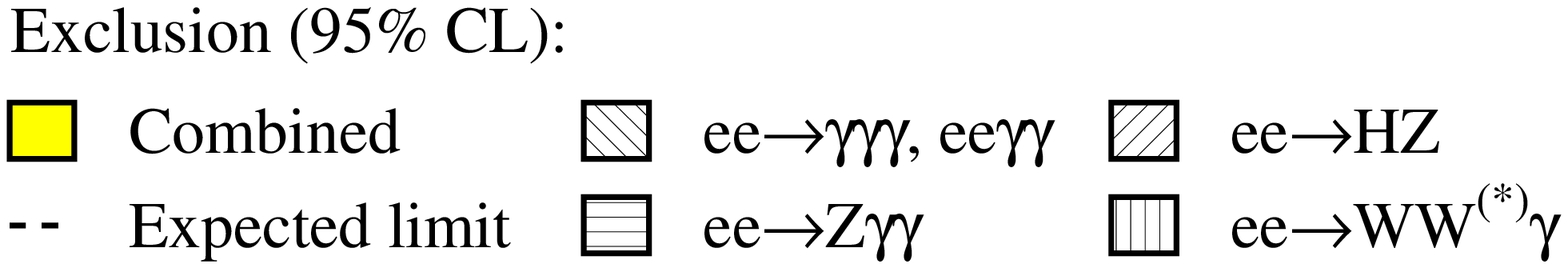} \\
	 \includegraphics[width=2.2in]{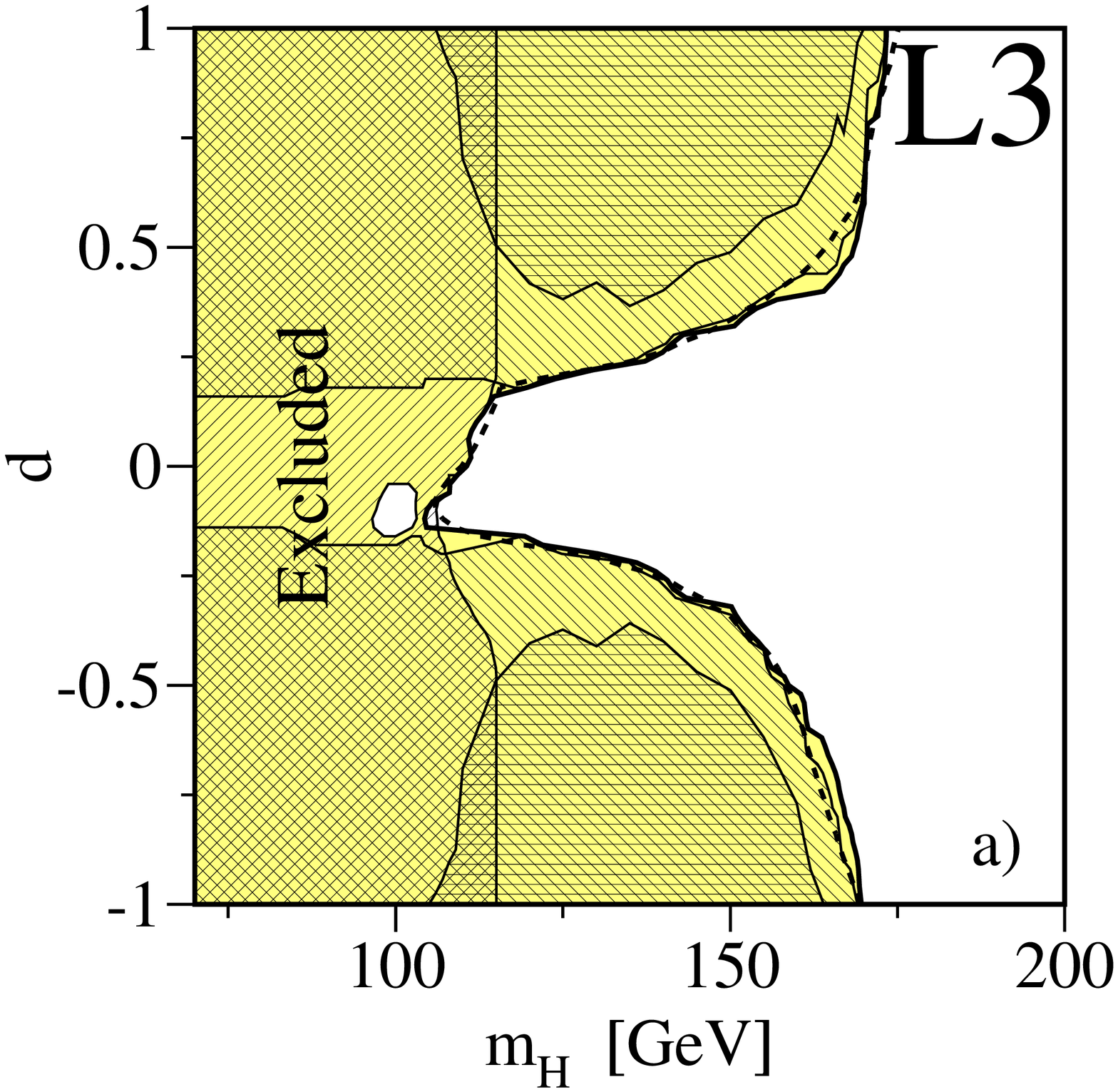} 
       \end{center}
       \vspace{-4.0mm}
       \caption{Regions excluded at 95\% CL for the anomalous coupling d.The different hatched regions show the limits obtained by the most sensitive analyses.}
       \label{fig7}
\vspace{-4.0mm}
     \end{figure}
      
     \section{Conclusions}
     The LEP experiments set very stringent limits on the existence of a Higgs in bSM scenarios.
     A large part of the accessible parameter space of such models could be excluded. This may help preparing the searches for the nature of the Higgs sector at future experiments, like those located at the LHC.  
          
     \balance

\end{document}